\documentstyle[aps,prb,preprint,tighten,eqsecnum]{revtex}

\begin{document}

\draft

\title{Quantum transport in disordered wires: Equivalence of one-dimensional
{\boldmath $\sigma$} model and Dorokhov-Mello-Pereyra-Kumar equation}

\preprint{cond-mat/9508056}

\author{P.\ W.\ Brouwer}

\address{Instituut-Lorentz, University of Leiden, P.O. Box 9506, 2300 RA
	Leiden, The Netherlands}

\author{K.\ Frahm}
\address{Service de Physique de l'\'Etat Condens\'e,
        Commissariat \`a l'Energie Atomique Saclay, 91191 Gif-sur-Yvette,
	France}

\date{August 15, 1995}

\maketitle

\begin{abstract}%
The two known non-perturbative theories of localization in disordered wires,
the Fokker-Planck approach due to Dorokhov, Mello, Pereyra, and Kumar,
and the field-theoretic approach due to Efetov and Larkin, are
shown to be equivalent for all symmetry classes. The equivalence
had been questioned as a result of field-theoretic calculations of
the average conductance by Zirnbauer [PRL 69, 1584 (1992)], which
disagreed with the Fokker-Planck approach in the symplectic symmetry
class. We resolve this controversy by pointing to an incorrect
implementation of Kramers degeneracy in these calculations,
and we derive modified expressions for the first two conductance moments
which agree well with existing numerical simulations from the metallic into
the localized regime. \bigskip
\end{abstract}%



\section{Introduction}
\label{sec:1}

There are two known approaches to the theory of phase-coherent conduction
and localization in disordered wires: The first is the
Fokker-Planck approach of Dorokhov, Mello, Pereyra, and
Kumar.\cite{dorokhov,mello2,mello3,mello4,macedo}
The second is the field-theoretic approach of Efetov and Larkin, which leads to
a supersymmetric nonlinear $\sigma$ model.\cite{larkin,efetov}
Both approaches provide a description of quantum transport that is independent
of microscopic details of the disordered wire. The only properties which enter
are its length $L$, the elastic mean free path $\ell$, the number $N$
of propagating transverse modes at the Fermi level (referred to as
``channels''), and the symmetry index $\beta \in \{1,2,4\}$
(depending on the presence or absence of time-reversal and/or spin-rotational
symmetry).  In the first approach, the transfer matrix is expressed as a
product of a large number of random matrices. As more matrices are added to
this product, the transmission eigenvalues $T_n$ execute a Brownian motion.
(The $T_n$ are the $N$ eigenvalues of the transmission matrix product
$t^{\dagger} t$.) The resulting Fokker-Planck equation for the $L$-dependence
of the distribution $P(T_1,\ldots,T_N)$ is known as the
Dorokhov-Mello-Pereyra-Kumar (DMPK) equation. In the second approach, one
starts
from the random Hamiltonian of the disordered wire and then expresses averages
of Green's functions\cite{larkin,efetov} or moments of the transmission
eigenvalues\cite{vwz,iwz,zirn1,mmz} as integrals over matrices $Q$
containing both
commuting and anticommuting variables. These so-called supermatrices are
restricted by the nonlinear constraint $Q^2=1$ and give rise to a
field theory known as the one-dimensional nonlinear $\sigma$ model.

In the last decade, research on the Fokker-Planck and field-theoretic approach
has proceeded quite independently. Recently, exact results for the average
conductance $\langle G \rangle$, its variance $\mbox{var}\, G$, and the
density $\rho(T) = \langle \sum_n \delta(T - T_n) \rangle$ of transmission
eigenvalues were obtained from both approaches. For the unitary symmetry
class (no time-reversal symmetry; $\beta=2$),
the DMPK equation was solved exactly by Beenakker and Rejaei.\cite{been1} The
construction of a set of biorthogonal polynomials for
this exact solution then allowed for the exact computation of
$\langle G \rangle$, $\mbox{var}\, G$, and $\rho(T)$ for arbitrary
$N$ and $L$ in the case $\beta=2$.\cite{frahm1} Although there exists
a formal solution for the other two symmetry classes [orthogonal class
(time-reversal symmetry without spin-orbit scattering; $\beta=1$) and
symplectic class (time-reversal symmetry with spin-orbit scattering;
$\beta=4$)],\cite{caselle} no exact results for $\langle G\rangle$,
$\mbox{var}\, G$, and $\rho(T)$ have been obtained.
Concerning the $\sigma$ model, an important and substantial progress
was the development of ``super Fourier analysis'' by Zirnbauer.\cite{zirn1}
This allowed the exact calculation\cite{zirn1,mmz} of $\langle G \rangle$
and $\mbox{var}\, G$ for all $\beta$ in the thick-wire limit
$N \rightarrow \infty$, $L/\ell \to \infty$ at fixed ratio $N \ell/L$.
The eigenvalue density $\rho(T)$ was computed
from the $\sigma$ model by Rejaei,\cite{rejaei} in the thick-wire limit and for
the case $\beta=2$.

If one takes the thick-wire limit of the $\beta=2$ results for
$\langle G \rangle$, $\mbox{var}\, G$, and $\rho(T)$ from the DMPK equation,
they agree precisely with those from the $\sigma$ model.\cite{frahm1,rejaei}
For $\beta=1$ and $4$, a comparison of the two approaches has only been
possible in the metallic regime $\ell \ll L \ll N \ell$, where the results
for $\langle G \rangle$ and $\mbox{var}\, G$
from the DMPK equation\cite{mello3,mello4,macedo} and from the
$\sigma$ model\cite{iwz,mmz,altland1} agree with conventional
diagrammatic perturbation theory.\cite{anderson,gorkov,altshuler,lee}
The equivalence of the two approaches outside the perturbative regime has
been questioned \cite{frahm1}
as a result of recent work by Zirnbauer,\cite{zirn1} and
by Mirlin, M\"uller-Groeling, and Zirnbauer.\cite{mmz} Starting from the
$\sigma$ model in the thick-wire limit, they obtained a finite limit
$\langle G \rangle \to e^2/2h$ as $L/N\ell \to \infty$ in the case $\beta=4$.
On the other hand, one can prove rigorously\cite{frahm1} that the DMPK equation
gives $\lim_{L \to \infty} \langle G \rangle = 0$ for all $\beta$.
It was this puzzling contradiction which motivated us to search for a general
proof of equivalence of the DMPK equation and the $\sigma$ model, without
the restriction to $\beta=2$.

In this paper, we present a general proof of the equivalence of the two
approaches,
which applies to all three symmetry classes $\beta$, to all length scales
$L$, and to the complete distribution of transmission eigenvalues described
by the
$p$-point functions $\rho_p(T_1,\ldots,T_p) =
(N!/(N-p)!) \int dT_{p+1}\ldots\int dT_N\ P(T_1,\ldots,T_N)$ for arbitrary $p$.
We cannot relax the assumption that the number $N$ of propagating
channels in the disordered wire is $\gg 1$, since it is needed for the
derivation of the one-dimensional $\sigma$ model.\cite{mmz} However, we
can consider the $\sigma$ model formulation of a thick disordered wire which
is coupled to the leads by means of a point contact with $N_1 \le N$
transmitted modes,\cite{iwz} and show that it is mathematically equivalent
to a DMPK equation for a wire with $N_1$ propagating channels.
The equivalence
proof demonstrates that $\lim_{L \to \infty} \langle G \rangle = 0$
in the $\sigma$ model, in apparent contradiction with Zirnbauer's work.
We have reexamined the calculation of Refs.\ \ref{zirn1} and \ref{mmz}, and
argue that for $\beta=4$ the Kramers degeneracy of the transmission eigenvalues
was not taken into account properly in the super Fourier analysis. This
leads to a spurious ``zero-mode'', which does not decay as $L \to \infty$.
Restoring Kramers degeneracy, we obtain modified
expressions for $\langle G \rangle$ and var$\,G$ which decrease
exponentially in the localized regime and moreover agree well with numerical
simulations.\cite{numerics}

Both the $\sigma$ model and the DMPK equation were derived from a number of
different models for a disordered wire.
The original derivation of the DMPK equation by Dorokhov,\cite{dorokhov}
which started from a model of $N$ coupled chains with defects, was followed
by the random-matrix formulation of Mello, Pereyra, and Kumar.\cite{mello2}
These authors considered a product of random transfer matrices, drawn from an
ensemble of maximum entropy. Later it was shown that the DMPK
equation is insensitive to the choice of the ensemble, the only relevant
assumptions being weak scattering (mean free path $\ell$ much greater than
the Fermi wave length $\lambda_F$) and equivalence of the scattering
channels.\cite{mello5,mello6} It is this latter assumption which restricts
the DMPK equation to a wire geometry.
{}From the mathematical point of view, the DMPK equation is the diffusion
equation on a certain coset-space of transfer matrices.\cite{hueffmann}
The one-dimensional $\sigma$ model was originally derived by Efetov and
Larkin\cite{larkin,efetov} from a white noise model for the disorder potential.
Two later derivations used random-matrix models for the
Hamiltonian of the disordered wire.
Iida, Weidenm\"uller, and Zuk (IWZ) adapted Wegner's $n$-orbital
model\cite{wegner}
to the study of transport properties.\cite{iwz} In this description, the wire
is modeled by a large number of disordered segments in series, each segment
having a random Hamiltonian drawn from the Gaussian ensemble. An alternative
derivation of the $\sigma$ model, due to Fyodorov and Mirlin,\cite{fyodorov}
uses a random band matrix to model the Hamiltonian of the disordered
wire. In the present paper we follow Ref.\ \ref{mmz} and use the IWZ
formulation of the $\sigma$ model.

Our proof of equivalence of the DMPK equation and the $\sigma$ model builds
on the ideas which were used by Rejaei\cite{rejaei} to
calculate $\rho(T)$ from the $\sigma$ model for $\beta=2$.
Inspired by Nazarov's diagrammatic calculation of $\rho(T)$ in the
metallic regime,\cite{nazarov} Rejaei introduced a generating
function $F$ which depends both on the transmission eigenvalues $T_n$
and on the radial parameters $\theta_i$ of the supermatrices in the
unitary $\sigma$ model. Rejaei was able to solve the $1d$ $\sigma$ model
exactly for $\beta=2$ and thus obtained the density $\rho(T)$ as a function of
$L$ by taking derivatives of $F$ with respect to the $\theta_i$'s. The
resulting $\rho(T)$ could then be compared with the result from the DPMK
equation.\cite{frahm1}
We introduce a more general generating function which allows us to establish
the equivalence of the $\sigma$ model and the DMPK equation at the level of
$p$-point functions $\rho_p(T_1,\ldots,T_p)$, without actually having to
compute this function. This approach works also for $\beta=1$ and $4$, where
no explicit solution of the $\sigma$ model is available.

The outline of the paper is as follows: In Sec.\ \ref{sec:2}, an outline
of the equivalence proof is given. The full proof for
the $\sigma$ model with $8\times 8$ supermatrices
follows in Secs.\ \ref{sec:3} and \ref{sec:4}, with technical material
in Apps.\ \ref{app:b} -- \ref{app:c}. For the $p$-point functions
$\rho_p(T_1,\ldots,T_p)$, we have to consider the $\sigma$ model with
$8p \times 8p$ supermatrices. This extension is described in App.\ \ref{app:d}.
In section \ref{sec:5}, we discuss the symplectic symmetry class ($\beta=4$)
in relation to Refs.\ \ref{zirn1} and \ref{mmz}. By accounting for
Kramers degeneracy, we obtain modified expressions for $\langle G \rangle$
and $\mbox{var}\, G$, which we compare with numerical simulations
of the IWZ model by Mirlin and M\"uller-Groeling.\cite{numerics}
We conclude in Sec.\ \ref{sec:6}.

\section{Outline of the equivalence proof} \label{sec:2}

Although our equivalence proof is technically rather involved, the basic
idea can be described in a few paragraphs.
In this section, we present an outline of the equivalence proof
for the small $\sigma$ model ($8 \times 8$ supermatrices).
The details are given in the following two sections and in the
appendices \ref{app:b} --- \ref{app:c}.
Appendix \ref{app:d} contains the necessary modifications to extend
the proof to $\sigma$ models with supermatrices of arbitrary size.

Part of the complexity of the problem is that the $\sigma$ model
and the DMPK equation focus on totally different objects.
In the $\sigma$ model, transport properties are expressed as functional
integrals over supermatrices $Q$.\cite{iwz,mmz} (A supermatrix is a matrix
containing an equal number of commuting and anticommuting elements. We
follow the notation and conventions of Refs.\ \ref{vwz}, \ref{iwz}, and
\ref{mmz}.) For the small $\sigma$ model the $8 \times 8$ supermatrices
are parameterized as\cite{efetov,vwz}
\begin{mathletters} \label{eq:Sigma}
\begin{eqnarray} \label{eq:6}
  && Q = T^{-1}\Lambda T, \quad
  \Lambda=\left(\begin{array}{cc}
   1 & \phantom{-}0 \\
   0 & -1 \\
  \end{array}\right), \\ \label{eq:Sigmab}
  && T=\left(
   \begin{array}{cc} u^{-1}\! & 0 \\ 0 & v^{-1} \\ \end{array}
  \right)\, \exp \left(
   \begin{array}{cc} 0 & \frac{1}{2}\hat\theta \\ \frac{1}{2}\hat\theta & 0 \\
  \end{array}\right) \left(
   \begin{array}{cc} u\, & 0 \\ 0 & v \\ \end{array}\right),
\end{eqnarray}
where $u$ and $v$ are pseudo-unitary $4 \times 4$ supermatrices. Notice
that $Q$ satisfies the non-linear constraint $Q^2=1$, hence the name
``non-linear'' $\sigma$ model. (The letter $\sigma$ is used for historical
reasons.)
The $4 \times 4$ supermatrix $\hat\theta$ is called the radial part of $Q$.
It has the form
\begin{equation}
  \label{eq:9}
  \hat\theta=\left(\begin{array}{cccc}
  \theta_1 & \theta_2 & 0 &  0 \\ \theta_2 & \theta_1 & 0 & 0 \\
  0 & 0 & i\theta_3  & i\theta_4 \\ 0 & 0 & i\theta_4  & i\theta_3
  \end{array}\right),
\end{equation}
with the symmetry restrictions
\begin{eqnarray}
\nonumber
&& \lefteqn{\theta_4 = 0} \hphantom{\theta_2=\theta_4 = 0}
 \ \ \mbox{if $\beta=1$},\\
\label{eq:10}
&& \theta_2=\theta_4 = 0 \ \ \mbox{if $\beta=2$},\\
\nonumber
&& \lefteqn{\theta_2 = 0} \hphantom{\theta_2=\theta_4 = 0}
 \ \ \mbox{if $\beta=4$}.
\end{eqnarray}
\end{mathletters}%

While the $\sigma$ model works with the radial part of a supermatrix,
the DMPK equation works with the radial part of an ordinary matrix
(containing only commuting elements). This is the transfer matrix $X$.
The radial part of $X$ is an $N \times N$ diagonal matrix $\hat \lambda$,
related to the eigenvalues of $X X^{\dagger}$. The eigenvalues of
$X X^{\dagger}$ come in $N$ inverse pairs $e^{\pm x_n}$, related to the
diagonal elements $\lambda_n$ of $\hat \lambda$ by $\lambda_n = \sinh^2 x_n$.
For $\beta=4$ the eigenvalues are twofold degenerate (Kramers degeneracy).
The matrix $\hat \lambda$ then contains only the $N$ independent eigenvalues.
The conductance $G$ is directly related to the $\lambda_n$'s by the Landauer
formula\cite{mello2,review_matrix}
\begin{equation}
  G = {2e^2 \over h} \sum_{n=1}^{N} T_n
    = {2 e^2 \over h} \sum_{n=1}^{N} {1 \over 1 + \lambda_n},
\end{equation}
since the $N$ independent transmission eigenvalues $T_n$ are related to the
$\lambda_n$'s by $T_n = (1+\lambda_n)^{-1}$.

We connect both approaches by considering a generating
function $F(\hat\theta,\hat\lambda)$ which depends on both radial matrices:
\begin{mathletters} \label{eq:1}
\begin{eqnarray} \label{eq:1a} \label{eq:23}
&&  F(\hat\theta,\hat\lambda) =
\prod_{n=1}^{N} f(\hat\theta,\lambda_n),
     \\
&& f(\hat\theta,\lambda) =
    \mbox{Sdet}^{-d/2} \left[\lambda
+ \cosh^2(\hat\theta/2)\right] =
\label{eq:22b}
\\
&& \left[
{\biglb(1 + 2 \lambda + \cos(\theta_3+\theta_4)\bigrb)\,
 \biglb(1 + 2 \lambda + \cos(\theta_3-\theta_4)\bigrb) \over
 \biglb(1 + 2 \lambda + \cosh(\theta_1+\theta_2)\bigrb)\,
 \biglb(1 + 2 \lambda + \cosh(\theta_1-\theta_2)\bigrb)}\right]^{{d / 2}},
  \nonumber\\
&&\nonumber\\
&& d=1\ \mbox{if $\beta=1,2$};\ \ \ d=2\ \mbox{if $\beta=4$}.
 \label{eq:dbetadef}
\end{eqnarray}
\end{mathletters}%
The symbol $\mbox{Sdet}$ stands for the superdeterminant of a supermatrix.
For $\beta=2$ this is the generating function introduced by Rejaei.

An ensemble of disordered wires of length $L$ provides a distribution of
$\hat\lambda$. The ensemble average $\langle F(\hat\theta,\hat\lambda)
\rangle$ contains all statistical properties that are accessible from the
small $\sigma$ model. These include the average conductance
$\langle G \rangle$, its variance $\mbox{var}\,G$ and the density of
transmission eigenvalues $\rho(T)$. We explain in appendix \ref{app:b}
how to extract these quantities by taking derivatives of
$\langle F(\hat\theta,\hat\lambda) \rangle$.
The average $\langle F(\hat\theta,\hat\lambda) \rangle$ can be determined
by each of the two approaches independently, in terms of a partial
differential equation for the $L$-dependence and an initial condition at
$L=0$.
For the $\sigma$ model on the one hand, the evolution equation reads
\begin{mathletters}
\begin{equation} \label{eq:Evolution}
  {\partial \over \partial L} \left\langle
    F(\hat\theta,\hat\lambda)\right\rangle =
  {2 \over \xi} \Delta_{\hat\theta}
    \left\langle F(\hat\theta,\hat\lambda)\right\rangle,
\end{equation}
where $\Delta_{\hat\theta}$ is the (radial part of the) Laplacian on the
$\sigma$ model space, and where $\xi = \beta N \ell$ is the localization
length. The explicit form of $\Delta_{\hat\theta}$ is given by\cite{efetov}
\begin{equation}
\label{eq:22}
  \Delta_{\hat\theta} =
   {\beta \over 2 d} \sum_i J^{-1}(\hat\theta)\,{\partial \over \partial
\theta_i}\,
   J(\hat\theta)\,{\partial \over \partial \theta_i},
\end{equation}
where the sum runs over the independent coordinates $\theta_i$ [see Eq.\
(\ref{eq:10})] and $J(\hat\theta)$ is the integration measure for the
radial decomposition (\ref{eq:Sigma}),
\begin{eqnarray}
  \nonumber
  J(\theta_1,\theta_2,\theta_3) & = &
  \sinh \theta_1\,\sinh \theta_2\,\sin^3 \theta_3
    \prod_{s_1,s_2=\pm 1} \sinh^{-2}\biglb(
    {\case{1}{2}} (\theta_1+ s_1\theta_2+i s_2\theta_3)\bigrb)
  \ \ \mbox{if $\beta=1$},\\
  \label{c1} \label{c2} \label{c3}
  J(\theta_1,\theta_3) & = &
  \lefteqn{\sinh \theta_1\,\sin \theta_3
    \prod_{s_1=\pm 1} \sinh^{-2}\biglb(
    {\case{1}{2}} (\theta_1+i s_1\theta_3)\bigrb),}
  \hphantom{\sinh \theta_1\,\sinh \theta_2\,\sin^3\theta_3
    \prod_{s_1,s_2=\pm 1} \sinh^{-2}\biglb(
    {\case{1}{2}} (\theta_1+ s_1\theta_2+i s_2\theta_3)\bigrb)}
  \ \ \mbox{if $\beta=2$},\\
  \nonumber
  J(\theta_1,\theta_3,\theta_4) & = &
  \lefteqn{\sin \theta_3\, \sin \theta_4\, \sinh^3 \theta_1
    \prod_{s_1,s_2=\pm 1} \sinh^{-2}\biglb(
    {\case{1}{2}}(\theta_1+ i s_1\theta_3+i s_2\theta_4)\bigrb)}
  \hphantom{\sinh \theta_1\,\sinh \theta_2\,\sin^3\theta_3
    \prod_{s_1,s_2=\pm 1} \sinh^{-2}\biglb(
    {\case{1}{2}} (\theta_1+ s_1\theta_2+i s_2\theta_3)\bigrb)}
  \ \ \mbox{if $\beta=4$}.
\end{eqnarray}
\end{mathletters}%

The DMPK equation on the other hand, yields the evolution equation
\begin{mathletters}
\begin{equation}
  {\partial \over \partial L}
    \left\langle F(\hat\theta,\hat\lambda)\right\rangle =
  {2 \over \xi}
    \left\langle D_{\hat\lambda}\, F(\hat\theta,\hat\lambda)\right\rangle,
\end{equation}
where $D_{\hat\lambda}$ is a second order differential operator in the
parameters $\lambda_n$,
\begin{eqnarray}
  D_{\hat\lambda} &=& J^{-1}(\hat\lambda)\sum_{n=1}^{N}
  {\partial \over \partial \lambda_n}\,
  J(\hat\lambda)\,\lambda_n(1+\lambda_n)\,{\partial \over \partial \lambda_n}
, \label{eq:29} \\
  J(\hat\lambda) &=& \prod_{n>m} |\lambda_n-\lambda_m|^\beta.
\end{eqnarray}
\end{mathletters}%
The key ingredient of the equivalence proof is the identity
\begin{equation} \label{eq:CentralResult} \label{eq:31} \label{EQ:31}
   \Delta_{\hat\theta} F(\hat\theta,\hat\lambda) =
    D_{\hat\lambda}\, F(\hat\theta,\hat\lambda),
\end{equation}
which shows that the evolution with $L$ of $\langle F(\hat\theta,\hat\lambda)
\rangle$ is the same in both approaches.
Showing that the initial conditions at $L=0$ coincide as well, completes
the equivalence proof.

\section{One-dimensional {\boldmath $\sigma$} model}
\label{sec:3}

We begin the detailed exposition of the equivalence proof with a formulation
of the $\sigma$ model. As in Ref.\ \ref{mmz}, we use the formulation of
Iida-Weidenm\"uller-Zuk (IWZ).\cite{iwz}

\subsection{The IWZ model}

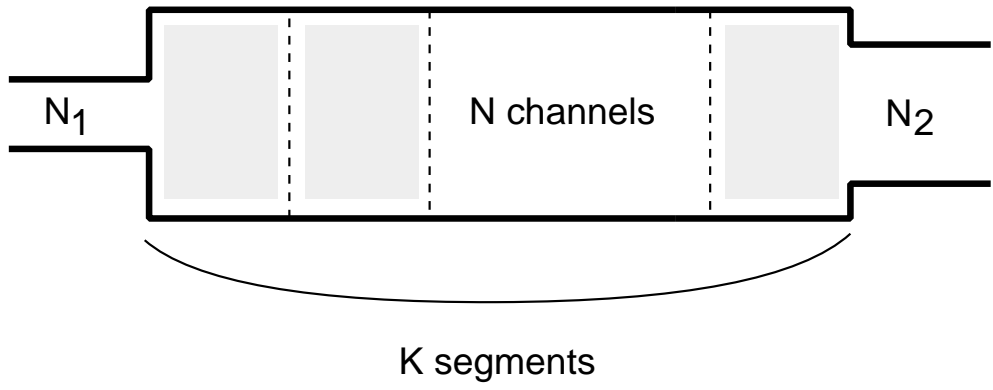
\begin{figure}
\caption{\label{figw} Schematic drawing of the disordered wire and
the leads according to the IWZ model description.
The left lead (lead $1$) contains $N_1$,
the right lead (lead $2$) $N_2$ propagating channels. The
number of propagating channels in the disordered wire is $N$. In the IWZ
model, the disordered wire is divided into $K$ segments, each having
a random Hamiltonian drawn drom the Gaussian ensemble.
The derivation of the $1d$ $\sigma$ model
from the IWZ model assumes $N \gg 1$, but allows for finite $N_1$ and $N_2$.}
\end{figure}

The IWZ model\cite{iwz,altland1} applies Wegner's $n$-orbital
model\cite{wegner} to a wire geometry and supplements it by a coupling
to ideal (not disordered) leads, as in Landauer's approach to
conduction.\cite{landauer}
The left and right leads (labeled by indices $1$ and $2$)
contain $N_1$ and $N_2$ propagating modes each (per spin direction
for $\beta=1,2$, or per Kramers doublet for $\beta=4$).
The disordered wire of length $L$ is assumed to consist of $K$ segments
in series (Fig.\ \ref{figw}).
The Hamiltonian $H$ of the disordered wire without leads is
represented by a matrix $H_{\mu \nu}^{ij}$, where the upper indices $i$, $j$
label the segments $1 \le i,j \le K$ and the lower indices $\mu$, $\nu$
label the $M$ states (per spin direction or Kramers doublet) within each
segment. The elements of $H$ are real ($\beta=1$), complex ($\beta=2$) or
quaternion ($\beta=4$) numbers. The coupling between the states inside one
segment is described by the matrices $H_{\mu\nu}^{ii}$, which are distributed
according to the Gaussian ensemble
\begin{equation}
  P(H^{ii}) = \mbox{const.}\, \times \,
  \exp\left(-\case{1}{4}\beta M v_1^{-2} \mbox{Tr}\, (H^{ii})^2\right).
  \label{GaussEns}
\end{equation}
Here $v_1$ is a parameter which governs the level density at the Fermi
level ($E=0$). The coupling between the states of adjacent segments is
given by another set of Gaussian distributed random matrices
$H^{ij}=(H^{ji})^\dagger$ (with coupling parameter $v_2$),
\begin{eqnarray}
  && P(H^{ij}) = \mbox{const.}\, \times \,
  \exp\left(-\case{1}{2}\beta M^2 v_2^{-2} \mbox{Tr}\, H^{ij} H^{ji}\right),
  \nonumber \\ &&
  j = i \pm 1.
\end{eqnarray}
Segments which are not adjacent are uncoupled, $H^{ij} = 0$ if $|i - j| \ge 2$.
The coupling to the ideal leads is described by a fixed $K M \times (N_1 +
N_2)$ rectangular matrix $W = W_1 + W_2$ with real ($\beta=1$), complex
($\beta=2$) or quaternion ($\beta=4$) elements. The matrix $W$ has
elements $W_{\mu n}^{i}$, where $i$ labels the segment, $\mu$ the states in
the segment, and $n$ the modes in the leads. The elements of $W_1$ (which
describes the coupling to lead $1$) are nonzero only for $i = 1$ and $1 \le n
\le N_1$; the elements of $W_2$ (coupling to lead $2$) are nonzero only for
$i = K$ and $N_1 < n \le N_1 + N_2$.

The scattering matrix $S$ (matrix elements $S_{nm}$) of the system at energy
$E$ is given by \cite{iwz}
\begin{equation}
\label{eq:3}
S=1-2\pi i W^\dagger (E-H+i\pi W W^\dagger)^{-1} W.
\end{equation}
The indices $n,m$ correspond to lead $1$ if $1 \le n,m \le N_1$ and to
lead $2$ if $N_1 < n,m \le N_1 + N_2$. The reflection and transmission matrices
$r, r', t, t'$ are submatrices of $S$,
\begin{equation} \label{eq:4}
S= \left(\begin{array}{cc} r & t' \\ t & r' \\ \end{array}\right).
\end{equation}
Since $S$ is unitary, the products $t^{\dagger} t$ and $t'^{\dagger} t'$
have the same set of non-zero eigenvalues, denoted by $T_n =
(1 + \lambda_n)^{-1}$. (If $N_2 > N_1$ there are also $N_2 - N_1$ transmission
eigenvalues which are zero, and can therefore be disregarded.)

\subsection{The generating function}

We now define the generating function $F(\hat\theta,\hat\lambda)$ introduced
in the previous section. We start from the the relationship (\ref{eq:3})
between the scattering matrix and the Hamiltonian in the IWZ model.
We consider the generating function
\begin{mathletters} \label{eq:GenerIWZ}
\begin{eqnarray} \label{eq:7}
  && F = \mbox{Sdet}^{-{1 \over 2}}(E-{\cal H}+i\pi W_1^{\vphantom{\dagger}}
W_1^\dagger Q +
         i\pi W_2^{\vphantom{\dagger}} W_2^\dagger \Lambda), \\ &&
  {\cal H} = H 1_{8}\ \ \mbox{if $\beta=1,4$};\ \ \
  {\cal H} = (\mbox{Re} H) 1_8 +
  i (\mbox{Im} H) \tau_3\ \ \mbox{if $\beta = 2$}.
\end{eqnarray}
\end{mathletters}%
Here $1_8$ is the $8 \times 8$ supersymmetric unit matrix and $\tau_3$ is
a diagonal matrix with elements $(1,-1,1,-1,1,-1,1,-1)$. The matrix $\Lambda$
was defined in Eq.\ (\ref{eq:6}). Note that $Q$ is an arbitrary supermatrix
as in Eq. (\ref{eq:Sigma}) and that it replaces the matrix $\Lambda$
in the coupling term of lead $1$.
In App.\ \ref{app:a} we show that $F$ depends only on the radial part $\hat
\theta$ of the matrix $Q$ and that the only
dependence on $H$ is through the transmission eigenvalues $T_n = (1 +
\lambda_n)^{-1}$. We also show that Eq.\ (\ref{eq:GenerIWZ}) reduces to the
function $F(\hat\theta,\hat\lambda)$ defined in Eq.\ (\ref{eq:1}) of the
previous section.

In the following, we evaluate the ensemble average $\langle F \rangle$
using the supersymmetric formalism. We first express $\langle F \rangle$
as a Gaussian
integral over an $8MK$-dimensional supervector $\psi$:
\begin{equation}
\label{eq:12c}
  \langle F \rangle = \left\langle
  \int{\cal D}\psi\, \exp\left(\case{1}{2}i\,
            \psi^\dagger\Lambda(E-{\cal H} +
            i \pi W_1^{\vphantom{\dagger}} W_1^\dagger Q +
            i \pi W_2^{\vphantom{\dagger}} W_2^{\dagger} \Lambda + i \epsilon
\Lambda)
            \psi\right) \right\rangle.
\end{equation}
The convergence of the Gaussian integral is assured by the parameterization
(\ref{eq:Sigma}) of the matrix $Q$.
Performing the standard steps, described in Refs.\ \ref{iwz} and \ref{mmz},
we obtain in the relevant limit $M\to\infty$
\begin{mathletters}
\begin{eqnarray}
\label{eq:13}
  \langle F \rangle
  &=& \int dQ_1\int dQ_K\ f_1(Q,Q_1)\ f_2(\Lambda,Q_K)\, W(Q_1,Q_K),\\
  \label{eq:14}
  W(Q_1,Q_K) &=& \int dQ_2\ldots \int dQ_{K-1}\ \exp\left(
  -\frac{d\,v_2^2}{2 v_1^2}\sum_{i=1}^{K-1} \mbox{Str}(Q_i\,Q_{i+1}),
\right) \\
\label{eq:15}
f_1(Q,Q_1) & = & \exp\left(-\frac{1}{2} d \sum_{n=1}^{N_1}
\mbox{Str}\,\ln(1+x_n\,Q Q_1)\right),\\
\label{eq:16}
f_2(Q,Q_K) & = & \exp\left(-\frac{1}{2}d \sum_{n=N_1+1}
^{N_1+N_2}
\mbox{Str}\,\ln(1+x_n\,Q Q_K)\right).
\end{eqnarray}
\end{mathletters}
The numbers $x_n$ denote the eigenvalues of the matrices $(\pi/v_1)
W_1^\dagger W_1^{\vphantom{\dagger}}$ (if $1\le n \le N_1$) or
$(\pi/v_1) W_2^\dagger W_2^{\vphantom{\dagger}}$ (if $N_1<n\le N_1+N_2$).
The integer $d$ was defined in Eq.\ (\ref{eq:dbetadef}).

Following Ref.\ \ref{mmz}, we consider the limit $v_1^2\ll v_2^2$.
Then the sum in (\ref{eq:14}) can be replaced by an integral and the
$Q_i$-integrals yield a path integral. The discrete number of segments $K$
becomes the continuous (dimensionless) variable $s$. The propagator
(\ref{eq:14}) can be identified with the heat kernel of the supersymmetric
space, determined by the heat equation\cite{mmz}
\begin{eqnarray}
\label{eq:18}
  2 \beta (v_2/v_1)^2 {\partial \over \partial s} W(Q',Q'') &=&
  \Delta_{Q'} W(Q',Q''), \nonumber \\
  \lim_{s\to 0} W(Q',Q'') &=& \delta(Q',Q'').
\end{eqnarray}
The precise definition of the Laplacian $\Delta_Q$ and the detailed
justification of Eq. (\ref{eq:18}) are contained in Ref.\ \ref{mmz}
($\Delta_Q$ in Eq. (\ref{eq:18}) differs by an additional factor $\beta/(8d)$
with respect to the notations of Ref.\ \ref{mmz}).
We thus arrive at the expression
\begin{equation}
\label{eq:19}
  \langle F \rangle =
    \int dQ'\int dQ''\,f_1(Q,Q')\,W(Q',Q'')\,f_2(\Lambda,Q'').
\end{equation}

The next step is to notice that $f_1(Q,Q')$ has the same symmetry
as the heat kernel, i.e.\ $f_1(T^{-1} Q T, T^{-1}Q' T)=f_1(Q,Q')$
where $T$ is an arbitrary element as described in (\ref{eq:6}). This
implies $\Delta_{Q'} f_1(Q,Q')=\Delta_{Q} f_1(Q,Q')$ and hence
$\langle F \rangle$ also satisfies the heat equation
\begin{equation}
\label{eq:20a} \label{eq:24a}
  2 \beta (v_2/v_1)^2 {\partial \over \partial s} \langle F \rangle =
  \Delta_Q \langle F \rangle.
\end{equation}
Since $\langle F \rangle$ only depends on the radial part $\hat\theta$ of
$Q$, it is sufficient to consider the radial part $\Delta_{\hat\theta}$ of
the Laplacian $\Delta_Q$. This radial part $\Delta_{\hat\theta}$ can be
written as in Eq.\ (\ref{eq:22}). We thus find that the ensemble average
$\langle F(\hat\theta,\hat\lambda) \rangle$ of the generating function
defined in Eq.\ (\ref{eq:1}) satisfies the partial differential equation
\begin{equation}
\label{eq:20} \label{eq:24}
 2 \beta (v_2/v_1)^2 {\partial \over \partial s}
   \langle F(\hat\theta,\hat\lambda) \rangle =
   \Delta_{\hat\theta} \langle F(\hat\theta,\hat\lambda) \rangle,
\end{equation}
with the initial condition implied by Eq.\ (\ref{eq:18}),
\begin{equation}
\label{eq:21a}
  \lim_{s \to 0}
  \left\langle F(\hat\theta,\hat\lambda) \right\rangle =
    \int dQ' f_1(Q,Q') f_2(\Lambda,Q').
\end{equation}
Together, Eqs.\ (\ref{eq:24}) and (\ref{eq:21a}) determine
the ensemble average of the generating function
$F(\hat\theta,\hat\lambda)$ evaluated in the framework of the nonlinear
$\sigma$ model.

The two limits of the IWZ model which were needed for the derivation of Eq.\
(\ref{eq:21a}), $M \to \infty$ and $v_1^2/v_2^2 \to 0$, restrict the
validity of Eq.\ (\ref{eq:21a}) to the case of weak disorder
($\ell \gg \lambda_F$) and thick wires ($N \gg 1$)
respectively.\cite{iwz,mmz}
Whereas the requirement of weak disorder is also needed for the DMPK equation,
the requirement that the number of channels in the disordered wire be large
is not. To see how the latter requirement follows from the condition
$v_1^2 \ll v_2^2$, we consider the expression for the average conductance
$\langle G \rangle$ in the diffusive metallic regime
($\ell \ll L \ll N \ell$),\cite{iwz,mmz}
\begin{equation} \label{GOhm}
  \langle G \rangle = {2e^2 \over h} {N \ell \over L} =
                      {2e^2 \over h} {4 v_2^2 \over v_1^2 s}.
\end{equation}
Taking the linear dimension of a segment of the disordered wire in the
IWZ model of order $\ell$ (i.e. $s\approx L/\ell$, see Ref.\ \ref{iwz}),
we find that $v_1^2 \ll v_2^2$
corresponds to $N \gg 1$. However, no restriction has been put to the numbers
$N_1$ and $N_2$ of propagating channels in the leads in the above
derivation of the $\sigma$ model, which allows us to consider finite values
of $N_1$ and $N_2$. This situation corresponds to the case in which
the thick disordered wire is coupled to the leads $1$ and $2$ by means of
point contacts, with $N_1$, $N_2$ open channels. As in Ref.\ \ref{mmz},
the case of a disordered wire without point contacts is recovered in the
limit $N_1, N_2 \to \infty$.

We conclude this section with some remarks about the choice of initial
conditions. In usual $\sigma$ model calculations,\cite{zirn1,mmz,rejaei}
one considers ideal coupling ($x_n=1$, $n=1,\ldots,N_1+N_2$) and identifies
$N = N_1 = N_2$ (equal number of channels in the leads and in the wire).
In the thick-wire limit $N \rightarrow \infty$ the function
$f_i(Q,Q')$ is just the delta function\cite{mmz} $\delta(Q,Q')$,
and $\langle F \rangle$ becomes identical to the heat kernel itself [cf.\
Eq.\ (\ref{eq:19})]:
\begin{equation}
\label{eq:21b}
  \langle F \rangle = W(Q,\Lambda),\ N_1 = N_2 = N \gg 1.
\end{equation}
For $\beta=2$, this result was derived by Rejaei.\cite{rejaei} In this
case $\langle F \rangle$ has the delta-function initial condition
$\lim_{s \to 0} \langle F \rangle = \delta(Q,\Lambda)$. To make contact with
the DMPK equation, we need a different ``ballistic'' initial condition,
such that all $T_n$'s are unity in the limit of zero wire length. To
achieve this, we take ideal coupling and assume that one of the leads has many
more channels than the other. To be specific, we fix $N_1$
and take the limit $N_2 \to \infty$. One then finds the initial condition
\begin{eqnarray}
\label{eq:35}
  && \lim_{s \to 0} \langle F \rangle =
\exp\left(-{\case{1}{2}} N_1 d\,\mbox{Str}\,
\ln(1+Q \Lambda)\right) \nonumber \\ && \ =
  \left(\frac{\cos \theta_3 + \cos \theta_4}
             {\cosh \theta_1 + \cosh \theta_2}
\right)^{N_1 d},\ 1 \le N_1 \ll N_2.
\end{eqnarray}
In the next section, we will see that this is precisely the ballistic
initial condition of the DMPK equation.

\section{DMPK equation}
\label{sec:4}

Let us now evaluate the ensemble average of the generating function
 (\ref{eq:1}) from the DMPK equation. The DMPK equation is a
Fokker-Planck-type equation for the $L$-evolution of the probability
distribution $P(\hat\lambda)$ of the
$\lambda_n$'s:\cite{dorokhov,mello2,mello3,mello4,macedo}
\begin{mathletters}
\begin{eqnarray}
\label{eq:25}
  && {1 \over 2}
  (\beta N + 2 - \beta) \ell {\partial \over \partial L} P(\hat\lambda) =
  \nonumber \\ && \ \ \ \
  \sum_{n=1}^N {\partial \over \partial \lambda_n}
  \lambda_n(1+\lambda_n) J(\hat\lambda)
  {\partial \over \partial \lambda_n} J^{-1}(\hat\lambda) P(\hat\lambda), \\
\label{eq:27}
  && J(\hat\lambda) = \prod_{n>m} |\lambda_n-\lambda_m|^\beta,
\end{eqnarray}
\end{mathletters}%
where $\ell$ denotes the mean free path in the disordered wire and $N$
the number of propagating modes. There is {\em no} restriction to $N \gg 1$
in the DMPK approach. We take the ballistic initial condition
\begin{equation}
  \lim_{L \to 0} P(\hat\lambda) = \prod_{n=1}^{N} \delta(\lambda_n - 0^{+}).
\end{equation}
The DMPK equation implies for $F(\hat\theta,\hat\lambda)$ the evolution
equation\cite{mello2,mello4}
\begin{eqnarray}
  {\partial \over \partial L} \langle F(\hat\theta,\hat\lambda)\rangle &=&
  {\partial \over \partial L} \int d\lambda_1 \ldots \int d\lambda_{N}
    F(\hat\theta,\hat\lambda)\,P(\hat\lambda)\
  \nonumber \\ &=&
  \frac{2}{\ell} (\beta N + 2 - \beta)^{-1}
    \left \langle D_{\hat\lambda}\, F(\hat\theta,\hat\lambda) \right\rangle,
\label{eq:28}
\end{eqnarray}
with the differential operator $D_{\hat\lambda}$ given by Eq.\ (\ref{eq:29}).
In Appendix \ref{app:c} we prove the algebraic identity between the two
different types of Laplacians (\ref{eq:22}) and (\ref{eq:29}) applied
to $F(\hat\theta,\hat\lambda)$,
\begin{equation} \label{eq:CentralResult2}
   \Delta_{\hat\theta} F(\hat\theta,\hat\lambda) =
    D_{\hat\lambda}\, F(\hat\theta,\hat\lambda).
\end{equation}
{}From Eqs.\ (\ref{eq:28}), and (\ref{eq:CentralResult2}) we conclude that
that the average $\langle F(\hat\theta,\hat\lambda)\rangle$,
calculated in the framework of the DMPK equation, also fulfills the evolution
equation (\ref{eq:24}) of the nonlinear $\sigma$ model, provided we
identify [cf.\ Eq.\ (\ref{GOhm})]
\begin{equation}
  {4 \over s} (v_2/v_1)^2 = {N \ell \over L} = {\xi \over \beta L},\ \ N \gg 1.
\end{equation}
Here we introduced the localization length $\xi = \beta N \ell$ (notice
that the definition of $\xi$ in Ref.\ \ref{mmz} differs by a factor $2/\beta$).

It remains to compare the initial conditions. The ballistic initial condition
for the DMPK equation implies
\begin{equation}
\label{eq:34}
  \lim_{L \to 0} \langle F(\hat\theta,\hat\lambda)\rangle =
  f(\hat\theta,\lambda = 0)^N =
  \left( \frac{\cos \theta_3 + \cos \theta_4}{\cosh
\theta_1+ \cosh\theta_2}
    \right)^{N d},
\end{equation}
which equals the initial condition (\ref{eq:35}) for the nonlinear
$\sigma$ model (The thick-wire limit $\lim_{L \to 0} \langle F \rangle =
\delta(Q,\Lambda)$ is obtained by letting $N \to \infty$
in the above expression). This proves the equivalence of both approaches,
as far as the generating function (\ref{eq:1}) is concerned. In Appendix
\ref{app:d} we extend the equivalence proof to $p$-point functions
$\rho_p(T_1,\ldots,T_p)$ for arbitrary $p$.

\section{The controversial symplectic ensemble}
\label{sec:5}

The main motivation of this work was to resolve a controversy between the
DMPK equation and the one-dimensional $\sigma$ model in the symplectic
symmetry class ($\beta=4$). On the one hand, the DMPK equation
implies\cite{frahm1} $\langle G \rangle \to 0$ as $L \to \infty$. On the
other hand, Zirnbauer\cite{zirn1} finds from the $\sigma$ model that
$\langle G \rangle \to \frac{1}{2} e^2/h$ as $L \to \infty$.

The equivalence proof presented in this paper has as a logical consequence
that $\langle G \rangle \to 0$ as $L \to \infty$ if $\langle G\rangle$ is
evaluated in the framework of the $\sigma$ model. To
demonstrate this, we apply the argument of Ref.\ \ref{frahm1}. The DMPK
equation implies for the average dimensionless conductance
$g = \sum_{n} (1+\lambda_n)^{-1}$ the evolution equation \cite{mello4}
\begin{equation} \label{eq:41}
  \xi {\partial \langle g \rangle \over \partial L} =
  - \beta \langle g^2 \rangle  - (2-\beta) \langle g_2 \rangle,
\end{equation}
with $g_2 = \sum_{n} (1 + \lambda_n)^{-2}$.
This relation also follows from the evolution equation (\ref{eq:24})
of the $\sigma$ model (expanding the generating function for small
$\theta_i$ and applying the results of appendix \ref{app:b}).
Since $0 \le g_2 \le g^2$, we have
\begin{equation} \label{eq:42}
  \xi {\partial \langle g \rangle \over \partial L} \le
  -{1 \over 2} \beta \langle g^2 \rangle \le 0.
\end{equation}
We suppose that $\lim_{L \rightarrow \infty} \langle g \rangle$
exists. Since $\partial \langle g \rangle / \partial L \le 0$
[Eq.\ (\ref{eq:42})] this implies
$\lim_{L \to \infty} \partial \langle g \rangle/\partial L = 0$. Hence
$\lim_{L \to \infty} \langle g^2 \rangle = 0$ by Eq.\ (\ref{eq:42}).
Since $\langle g \rangle^2 \le \langle g^2 \rangle$ this implies that
also $\lim_{L \to \infty} \langle g \rangle = 0$.

Where does the non-zero limit in Refs.\ \ref{zirn1} and \ref{mmz} come
from? The ground-breaking contribution of Zirnbauer was to use a
``super-Fourier expansion'' of the heat kernel $W(Q,Q')$ in terms of
eigenfunctions of the
Laplacian in the space of the $\sigma$ model. This type of
Fourier analysis is well understood for classical symmetric
spaces.\cite{helgason} The development and application of the supersymmetric
analogue for the $\sigma$ model enabled
Zirnbauer, Mirlin, and M\"uller-Groeling
to compute non-perturbatively the first two moments of the conductance for any
$\beta$. The non-zero limiting value $\lim_{L \to \infty} \langle g \rangle =
1/4$ for $\beta=4$ resulted from a ``zero mode'', a non-trivial eigenfunction
of the Laplacian with zero eigenvalue. Since this zero mode does not decay
as $L \to \infty$, it led to the surprising conclusion of absence of
localization in a wire with spin-orbit scattering in zero magnetic
field.\cite{zirn1}

An explicit expression for the zero-mode was not obtained in Refs.\ \ref{zirn1}
and \ref{mmz}, but only its contribution to the moments of the conductance
was computed. By inspecting the initial condition (\ref{eq:35}) of the
generating function for the $\sigma$ model we have been able to construct
a zero mode for $\beta=4$, but only if we ignore the Kramers degeneracy of
the transmission eigenvalues. This unphysical zero mode, given by
\begin{equation}
\label{eq:zeromode}
   \phi_{0}(\theta_1,\theta_3,\theta_4) =
\frac{\cos\theta_3 + \cos\theta_4}{2 + 2\cosh \theta_1},
\end{equation}
arises by taking the initial condition (\ref{eq:35}) with $N_1 = 1$ and
$\beta=4$, but {\em without} the extra factor two in the exponent, required
by Kramers degeneracy.
This unphysical initial condition solves the evolution equation (\ref{eq:24})
for the ensemble average of the generating function and implies an
$L$-independent average conductance $\langle g \rangle = 1/4$. Although
we can not prove that Eq.\ (\ref{eq:zeromode}) is Zirnbauer's zero mode,
the coincidence with the limiting value $\lim_{L \to \infty} \langle g \rangle
= 1/4$, $\lim_{L \to \infty} \mbox{var}\, g = 1/16$ is quite suggestive.

The reason why we have to exclude the zero mode (\ref{eq:zeromode})
from the Fourier expansion of the heat kernel is that it is not
single-valued on the $\sigma$ model space of supermatrices $Q$,
although it is a well-defined function of $\hat\theta$. The parameterization
(\ref{eq:Sigma}) of $Q$ is $2\pi$-periodic in the angles
$\theta_\pm=\theta_3\pm\theta_4$. We can then consider on the space of
angles $\theta_3$, $\theta_4$ a parity operation $P$ which
consists of adding $\pi$ to both of these angles.
This parity operation does not change $Q$, but it changes the zero mode
(\ref{eq:zeromode}). The Laplacian (\ref{eq:22}) commutes with $P$ and
 the eigenfunctions have therefore either even or odd
parity (eigenvalues $+1$ or $-1$ of $P$, respectively). The
physical modes of the $\sigma$ model must have even parity, since only these
functions are single-valued. For $\beta=4$, it is the Kramers degeneracy
which ensures that the initial condition (\ref{eq:35}) has even parity.

This observation led us to check the parity of the eigenfunctions
$\phi_{\nu}(Q)$ of the Laplacian in the super Fourier analysis of Refs.\
\ref{zirn1} and \ref{mmz}. We consider the eigenvalue equation
\begin{equation}
  \Delta_{\hat\theta} \phi_\nu(\theta_1,\theta_3,\theta_4) =
  - \varepsilon(\nu) \phi_\nu(\theta_1,\theta_3,\theta_4)
\end{equation}
for $\beta=4$ in the limit $\theta_1 \rightarrow \infty$ at fixed
$\theta_3$, $\theta_4$. In this limit, the Laplace operator simplifies
considerably
\begin{equation}
\label{eq:43}
\Delta_{\hat\theta} \to
   e^{\theta_1}{\partial \over \partial \theta_1}
    e^{-\theta_1} {\partial \over \partial \theta_1} +
  {1 \over \sin \theta_3} {\partial \over \partial \theta_3}
    \sin \theta_3 {\partial \over \partial \theta_3} +
  {1 \over \sin \theta_4} {\partial \over \partial \theta_4}
    \sin \theta_4 {\partial \over \partial \theta_4}.
\end{equation}
{}From this expression one may identify the set of quantum numbers
$\nu=(\lambda,1+2n_1,1+2n_2),$ where $\lambda$ is a real number
and $n_1$, $n_2$ are
non-negative integers.
The asymptotic behavior of the eigenfunctions
$\phi_{\nu}(\theta_1,\theta_3,\theta_4)$ is given by
\begin{equation}
  \label{eq:42b}
  \phi_{\nu}(\theta_1,\theta_3,\theta_4)
  \sim \exp \left[{\case{1}{2}} (1+i\lambda)\theta_1\right]
\biggl(P_{n_1}(\cos\theta_3) P_{n_2}(\cos\theta_4) +
P_{n_2}(\cos\theta_3) P_{n_1}(\cos\theta_4)\biggr),
\end{equation}
with the Legendre polynomials $P_n(x)$ and the eigenvalues
\begin{equation} \label{eq:44}
  \varepsilon(\lambda,1+2n_1,1+2n_2) =
  \frac{1}{4}\biggl(\lambda^2 + (1+2n_1)^2 + (1+2n_2)^2 - 1\biggr).
\end{equation}
The parity of this eigenfunction is just $(-1)^{n_1+n_2}$ and we have to
restrict ourselves to those $n_1$ and $n_2$ with $n_1 + n_2$ even. Applying
this
selection rule to the expressions for $\langle g \rangle$ and $\langle g^2
\rangle$ of Refs.\ \ref{zirn1} and \ref{mmz}, omitting the zero mode [and
the subsidiary series with quantum numbers $\nu=(i,l,l\pm 2)$ of Refs.\
\ref{zirn1} and \ref{mmz}, for which the asymptotic behavior (\ref{eq:42b})
is also valid], and multiplying the surviving terms with a factor of $2$
to account for Kramers degeneracy, yields for $\beta=4$ and in the limit
$N_1 = N_2 = N \rightarrow \infty$ the expression
\begin{mathletters} \label{eq:49}
\begin{eqnarray}
  \langle g^n \rangle &=& 2^{5-n} \sum_{{l_1, l_2 = 1, 3, 5, \ldots,
\atop l_1 + l_2 \equiv
2\, (\mbox{\scriptsize mod}\, 4)}} \int_0^{\infty} d\lambda\
\lambda (\lambda^2
+ 1) \tanh(\pi\lambda/2) l_1 l_2 p_n(\lambda,l_1,l_2) \nonumber \\
&& \times
\prod_{\sigma,\sigma_1,\sigma_2 = \pm 1} (-1 + i \sigma \lambda +
\sigma_1 l_1
+ \sigma_2 l_2)^{-1} \exp\left[-(\lambda^2 + l_1^2 + l_2^2 -1) L / (2 \xi)
\right],
\end{eqnarray}
where $n = 1,2$ and
\begin{eqnarray}
  p_1(\lambda,l_1,l_2) &=& \lambda^2 + l_1^2 + l_2^2 -1, \\
  p_2(\lambda,l_1,l_2) &=& \case{1}{4} \biglb[2 \lambda^4 + l_1^4 + l_2^4 + 3
\lambda^2(l_1^2 + l_2^2) - 2 \lambda^2 + l_1^2 + l_2^2 - 2 \bigrb].
\end{eqnarray}
\end{mathletters}%
Note that in our notations the dimensionless conductance $g$ is by
a factor $2$ smaller than $g$ in the notations of Ref. \ref{mmz}.
Comparison of Eq.\ (\ref{eq:49}) with the $\beta=4$ result of Ref.\ \ref{mmz},
where the parity selection rule was not implemented, shows that the
perturbation expansion around $L/\xi = 0$ is the same. (We checked this
numerically up to order $(L/\xi)^3$.) Outside the perturbative regime, the
two expressions are completely different. Instead of a non-zero limit
$\langle g \rangle = 1/4$ for $L/\xi \gg 1$, we find from Eq.\ (\ref{eq:49})
the exponential decay
\begin{equation}
  \langle g \rangle \approx {16 \over 9}\, (2L/\pi\xi)^{-3/2} e^{-L /2\xi}.
\end{equation}

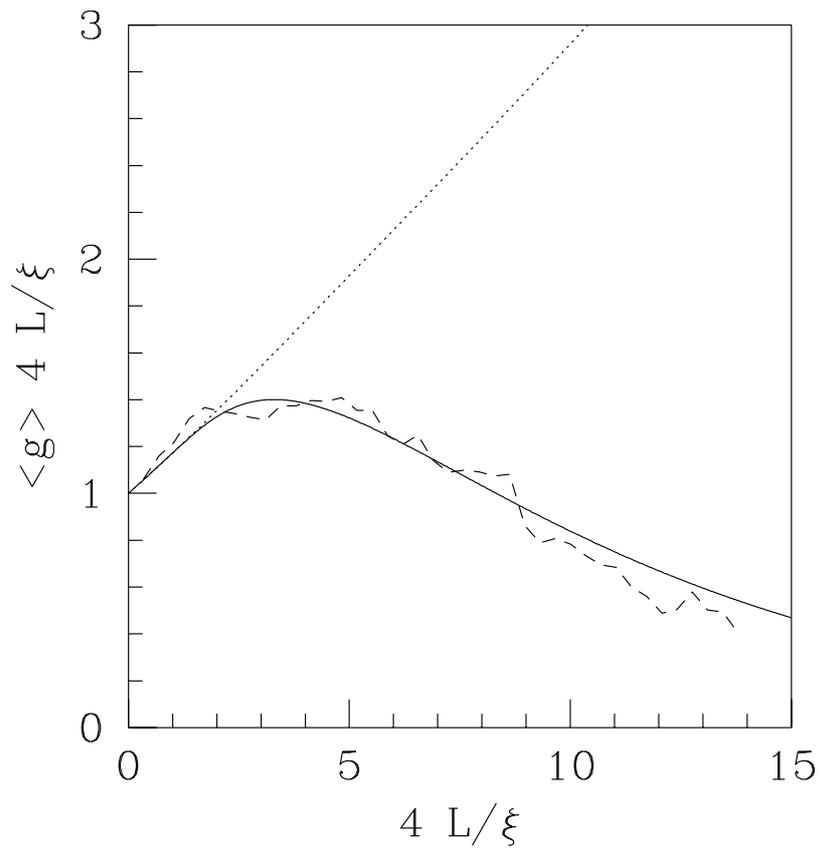
\begin{figure}
\caption{\label{fig1} The average conductance $\langle g\rangle$ multiplied
by $4L/\xi=L/N\ell$ for the symplectic symmetry class as a function
of $4L/\xi$ for $N \gg 1$. Shown are our result (\protect\ref{eq:49})
(solid), the numerical simulation of Ref.\ \protect \ref{numerics} ($M=100$,
$N=25$) (dashed), and the result of Ref.\ \protect\ref{mmz} (dotted).}
\end{figure}

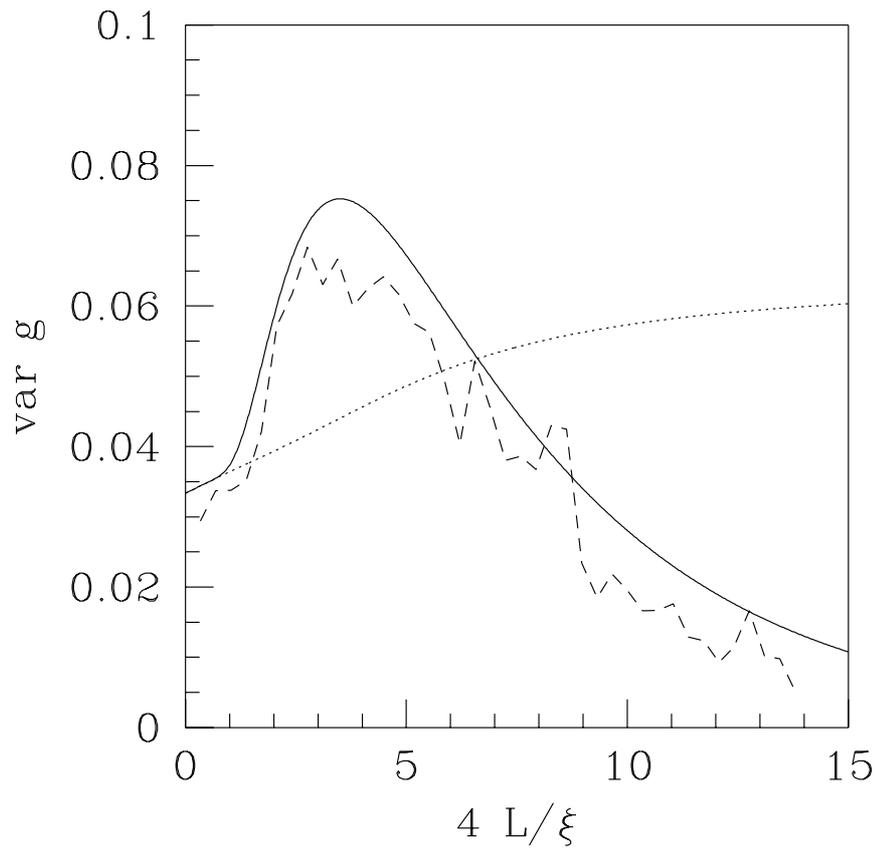
\begin{figure}
\caption{\label{fig2} As in Fig. \protect\ref{fig1} for the variance var$\,g$
of the conductance.}
\end{figure}

To test our result, we have compared it with a direct numerical simulation
of the IWZ model by Mirlin and M\"uller-Groeling\cite{numerics} (with
$M=100$, $N=25$ and an average over $100$ different samples). The comparison
is shown in Figs.\ \ref{fig1} and \ref{fig2}. It is clear that our
Eq.\ (\ref{eq:49}) (solid curve) agrees quite well with the simulation, while
the result of Ref.\ \ref{mmz} does not (dotted curve).

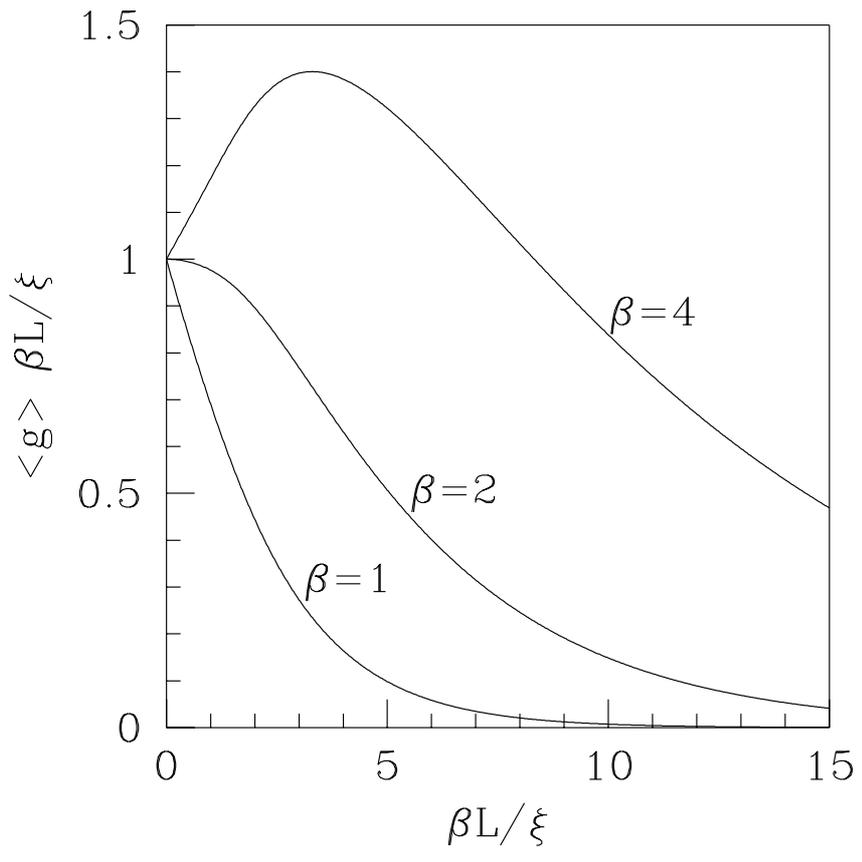
\begin{figure}
\caption{\label{fig3}The average conductance $\langle g\rangle$ multiplied by
$\beta L/\xi=L/N\ell$ for the three symmetry classes as a function
of $\beta L/\xi$ for $N \gg 1$. The curves for $\beta=1,2$ are taken from
Refs.\
\protect\ref{zirn1} and \protect\ref{mmz} and the curve for $\beta=4$
is calculated from Eq.\ (\protect\ref{eq:49}). Notice that $\xi = \beta N \ell$
is proportional to $\beta$, so that the scaling of the axes is
$\beta$-independent.}
\end{figure}

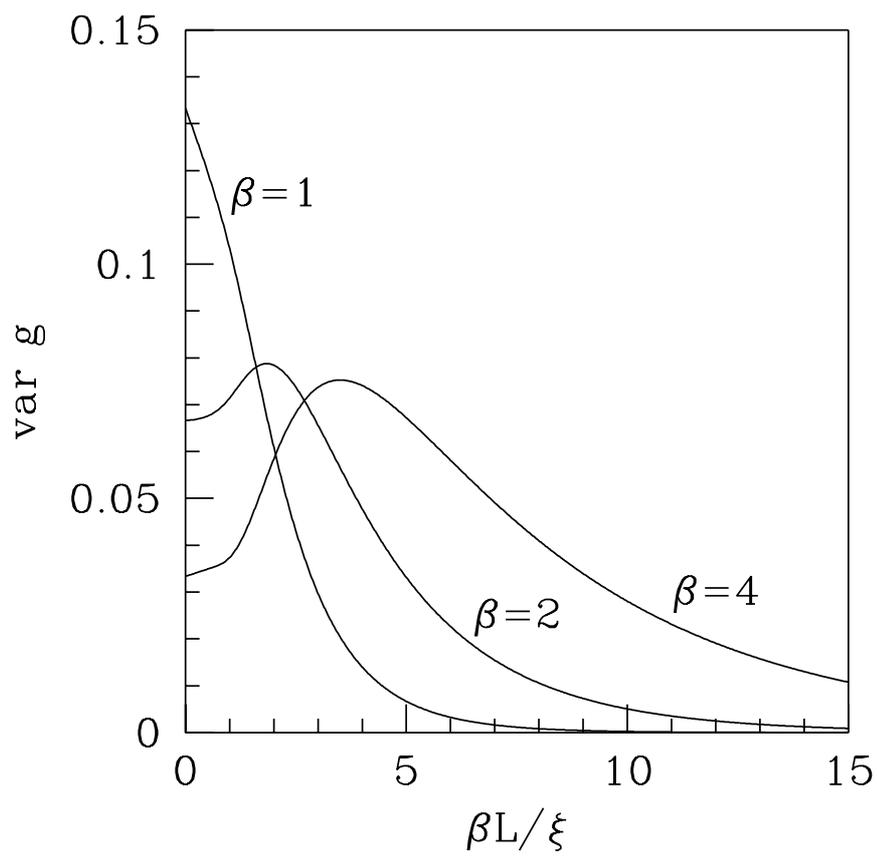
\begin{figure}
\caption{\label{fig4} Same as Fig.\ \protect\ref{fig3},
for the variance var$\,g$
  of the conductance}
\end{figure}

Notice that this issue of the parity of the eigenfunctions does not
occur for $\beta=1,2$, since there is only one compact angle ($\theta_3$)
in those cases. The parity operation on the $\hat\theta$-matrices exists
only for $\beta=4$. For completeness we collect in Figs.\ \ref{fig3} and
\ref{fig4} the results for $\langle g \rangle$ and $\mbox{var}\, g$ for
all three symmetry classes. The $\beta=1,2$ results are from Ref.\ \ref{mmz},
the $\beta=4$ result is our Eq.\ (\ref{eq:49}).

\section{Conclusion} \label{sec:6}

We have established the exact mathematical equivalence
of the two non-perturbative theoretical approaches to phase-coherent
transport and localization in disordered wires: The Fokker-Planck equation
of Dorokhov, Mello, Pereyra, and
Kumar\cite{dorokhov,mello2,mello3,mello4,macedo}
and the one-dimensional supersymmetric nonlinear $\sigma$
model.\cite{larkin,efetov,iwz,mmz,fyodorov}
The equivalence has the logical consequence that the absence of localization
in the symplectic symmetry class, obtained by Zirnbauer by super-Fourier
analysis of the $\sigma$ model, is not correct. By applying a selection rule
enforced by Kramers degeneracy to the eigenfunctions of Refs.\ \ref{zirn1}
and \ref{mmz}, we have obtained modified expressions for $\langle G \rangle$
and $\mbox{var}\, G$, which decay exponentially as $L \to \infty$ and which
agree well with existing numerical simulations.\cite{numerics}

Our equivalence proof has both conceptual and practical implications.
The DMPK equation and the $1d$ $\sigma$ model originated almost
simultaneously in the early eighties, and at the same institute.
\cite{dorokhov,larkin} Nevertheless, work on both approaches
proceeded independently in the next decade. Knowing that, instead
of two theories, there is only one, seems to us a considerable conceptual
simplification of the field. It implies that the microscopic derivations
and random-matrix models developed for the $\sigma$ model apply as well
to the DMPK equation, and vice versa. (we see only the restriction, that
the $\sigma$ model requires the thick-wire limit $N \to \infty$, while
the DMPK equation applies to any number of channels $N$.) Practically,
each of the two approaches has its own advantages, and now that we know
that they are equivalent, we can choose the approach which is best suited
to our needs and skills.

\acknowledgements
We thank C. W. J. Beenakker, B. Rejaei, and M. R. Zirnbauer
for fruitful discussions. We are very grateful to A. D. Mirlin and
A. M\"uller-Groeling for providing us with unpublished data from
their numerical simulation of the IWZ model.
This work was supported by the ``Stich\-ting voor
Fun\-da\-men\-teel On\-der\-zoek der Ma\-te\-rie'' (FOM) and by the
``Ne\-der\-land\-se or\-ga\-ni\-sa\-tie voor We\-ten\-schap\-pe\-lijk
On\-der\-zoek'' (NWO) (P.\ W.\ B.) and by the Human Capital Mobility
program of the European Community ``Quantum Dynamics of
Phase-Coherent Structures'' (K.\ F.).

\appendix

\section{Transport properties determined by the generating
function}
\label{app:b}

We list the transport properties of interest that can be generated from
$F(\hat\theta,\hat\lambda)$, following Rejaei.\cite{rejaei} Let us
consider the function
\begin{equation}
\label{eq:b1}
f(z_1,z_2)=\left\langle\frac{\det(1+z_2\,t^\dagger t)}
{\det(1+z_1\,t^\dagger t)}\right\rangle,
\end{equation}
which equals $\langle F(\hat\theta,\hat\lambda) \rangle$ at
$z_2=-\sin^2(\frac{1}{2}\theta_3)$, $z_1=\sinh^2(\frac{1}{2}\theta_1)$,
and $\theta_2=\theta_4=0$.
We write Eq.\ (\ref{eq:b1}) in the form
\begin{equation}
f(z_1,z_2)=\left\langle
\exp\left[\mbox{Tr}\,\ln(1+z_2\,t^\dagger t)-
\mbox{Tr}\,\ln(1+z_1\,t^\dagger t)\right]\right\rangle.
\end{equation}
The standard expansion \cite{vwz,iwz} with respect to small
$z_1$ and $z_2$ yields the first two moments of the
dimensionless conductance $g=(1/d)\,\mbox{Tr}\,t^\dagger t$
(with $d=1+\delta_{4,\beta}$),
\begin{eqnarray}
\langle g\rangle & = & {1 \over d} {\partial \over \partial z_2} f(z_1,z_2)
 \Big|_{z_1 = 0 = z_2} \nonumber \\ &=&
-{1 \over d} {\partial \over \partial z_1}f(z_1,z_2) \Big|_{z_1 = 0 = z_2},\\
\label{eq:b4}
\langle g^2\rangle & = & - {1 \over d^2} {\partial \over \partial z_1}
{\partial \over \partial z_2}
f(z_1,z_2)\Big|_{z_1 = 0 = z_2}.
\end{eqnarray}

We may also consider \cite{rejaei,nazarov} derivatives of $f(z_1,z_2)$ at
$z_1 = z_2$. This may require the analytic
continuation of $\theta_1$, $\theta_3$ to complex values if
$z_1<0$, $z_2>0$ or $z_2<-1$. Therefore, we introduce the
function $f(z_1)$ as
\begin{eqnarray}
\label{eq:b5}
  f(z_1)&=& {\partial \over \partial z_2}f(z_1,z_2) \Big|_{z_2 = z_1}
  \nonumber \\ & = & \left\langle\mbox{Tr}\,
\left[(1+z_1\, t^\dagger t)^{-1}\, t^\dagger t\right]
\right\rangle \nonumber \\ &=& \nonumber
\sum_{n=0}^\infty (-z_1)^{n} \langle \mbox{Tr}\, (t^{\dagger} t)^{n+1}
\rangle\\
& = & z_1^{-1}\biglb(\mbox{Tr}(1)-\left\langle\mbox{Tr}\,\left[(1+z_1\,
t^\dagger t)^{-1}\right]\right\rangle\bigrb).
\end{eqnarray}
The average density of transmission eigenvalues now follows from:
\begin{eqnarray}
\rho(T) &=& \langle\mbox{Tr}\,\delta(T-t^\dagger t)\rangle \nonumber \\ &=&
-\frac{1}{\pi T^2} \mbox{Im}
\ f\left(-(T+i 0^{+})^{-1}\right).
\label{eq:b6}
\end{eqnarray}
The application of Eq.\ (\ref{eq:b6}) requires the analytical continuation
of both variables $z_1$ and $z_2$ to values $<-1$.

\section{The generating function in terms of the transmission matrix}
\label{app:a}

In this appendix, we show that Eq.\ (\ref{eq:GenerIWZ}) for the generating
function
in the IWZ model equals Eq.\ (\ref{eq:1}). We first consider the two cases
$\beta=1,4$ of time reversal symmetry, when ${\cal H} = H 1_8$ in Eq.\
(\ref{eq:7}). The necessary modifications for $\beta=2$ are described at the
end.

We make use of the folding identity
\begin{equation}
\label{eq:a1}
\mbox{Sdet}
\left(\begin{array}{cc}
1_n & A   \\
B   & 1_m \\
\end{array}\right)=
\mbox{Sdet}(1_n-AB)=
\mbox{Sdet}(1_m-BA).
\end{equation}
We abbreviate $G_{\pm}=(E-H\pm i\pi W W^\dagger)^{-1}$. Taking out the factor
$(E-{\cal H}+i\pi W W^\dagger \Lambda)$ (with unit superdeterminant),
we may rewrite Eq.\ (\ref{eq:7}) as
\begin{eqnarray}
\nonumber
 F &=&
  \mbox{Sdet}^{-{1 \over 2}}\left(1+ \left(\begin{array}{cc}
  G_+ &  0 \\ 0 & G_- \\
  \end{array}\right)
  i\pi W_1^{\vphantom{\dagger}} W_1^\dagger(Q-\Lambda)\right)\\
\label{eq:a3}
&=& \mbox{Sdet}^{-{1\over 2}}\left(1+i\pi \left(\begin{array}{cc}
  W_1^\dagger G_+ W_1^{\vphantom{\dagger}} & 0 \\ 0 & W_1^\dagger G_-
W_1^{\vphantom{\dagger}} \\
\end{array}\right) (Q-\Lambda)\right),
\end{eqnarray}
where we have applied Eq.\ (\ref{eq:a1}) on $B=W_1^\dagger$. We now
insert the reflection matrix
$r = 1 - 2\pi i W_1^\dagger G_+ W_1^{\vphantom{\dagger}}$ [see
Eqs.\ (\ref{eq:4}) and (\ref{eq:3})] into Eq.\ (\ref{eq:a3}) and obtain
\begin{equation}
\label{eq:a4}
  F =
  \mbox{Sdet}^{-{1\over 2}}\left(\frac{1}{2}(1+\Lambda Q)+\frac{1}{2}
  \left(\begin{array}{cc}
  r & 0 \\ 0 & r^\dagger \\
\end{array}\right)(1-\Lambda Q)\right).
\end{equation}
Now we use the parameterization (\ref{eq:Sigma}) for $Q$. Notice that
Eq.\ (\ref{eq:a4}) does not depend on the angular part
of $Q$ [the matrices $u,v$ in (\ref{eq:Sigmab})]. Hence we may choose $Q$ as
\begin{equation}
\label{eq:a5}
Q=T^{-1}\Lambda T,\ T=\exp({\case{1}{2}}\Theta)
,\ \Theta=
\left(\begin{array}{cc}
 0 & \ \hat\theta \\
\hat\theta &  0 \\
\end{array}\right),
\end{equation}
which leads to
$$
{\textstyle
\frac{1}{2}(1+\Lambda Q)=\cosh(\frac{1}{2}\Theta)\,T,\ \
\frac{1}{2}(1-\Lambda Q)=-\sinh(\frac{1}{2}\Theta)\,T.
}
$$
Inserting this in Eq.\ (\ref{eq:a4}) and taking out the factor $T$
(with unit superdeterminant), we get
\begin{eqnarray}
\nonumber
  F & = &
  \mbox{Sdet}^{-1/2} \left(\begin{array}{cc}
  \cosh(\frac{1}{2_{\vphantom{M}}}\hat\theta)_{\vphantom{M}}
   & -r\sinh(\frac{1^{\vphantom{M}}}{2}\hat\theta) \\
  -r^\dagger\sinh(\frac{1}{2}\hat\theta)^{\vphantom{M}} &
  \cosh(\frac{1}{2}\hat\theta) \\
  \end{array}\right)\\
\nonumber &&\\
\nonumber
  & = & \mbox{Sdet}^{-1/2}
\left(\cosh^2(\case{1}{2}\hat\theta)-\sinh^2(\case{1}{2}\hat\theta)
r^\dagger r \right) \\
\label{eq:a7}
 & = & \mbox{Sdet}^{-1/2}
 \left(1+\sinh^2(\case{1}{2}\hat\theta) t^\dagger t\right),
\end{eqnarray}
where we have again used (\ref{eq:a1}) and the relation
$r^\dagger r=1- t^\dagger t$ imposed by unitarity of the scattering matrix.

The matrix $t^{\dagger} t$ has eigenvalues $T_n=(1+\lambda_n)^{-1}$
($n = 1, \ldots,N_1$), which are twofold degenerate for $\beta=4$, hence
\begin{eqnarray}
  \nonumber
  F & = &
  \prod_{n=1}^{N_1} \mbox{Sdet}^{-d/2}
 {\textstyle (1+\sinh^2(\frac{1}{2}\hat\theta)\, T_n)} \\
\label{eq:a8}
  & = & \prod_{n=1}^{N_1} \mbox{Sdet}^{-d/2} \left(\lambda_n
+ \cosh^2(\case{1}{2}\hat\theta)\right).
\end{eqnarray}
Here $d = 1 + \delta_{\beta,4}$. Eq. (\ref{eq:a8}) establishes
the connection between the two expressions (\ref{eq:1}) and
(\ref{eq:GenerIWZ}) for the generating function.

The above calculation assumes ${\cal H} = H 1_8$, hence $\beta=1$ or $\beta=4$.
In the case $\beta=2$, one has instead ${\cal H}=\mbox{Re}(H)
1_8+i\,\mbox{Im}(H)\tau_3$.
Instead of two subblocks with $r$ and $r^\dagger$
[see Eq. (\ref{eq:a4})], one now needs four subblocks with $r$, $r^T$,
$r^\dagger$, and $r^*$. Repeating the calculation one finds that the final
result (\ref{eq:a8}) remains the same.

\section{Identity of Laplacians}
\label{app:c}

The goal of this appendix is to prove Eq.\ (\ref{eq:31}). Hereto we
first analyze the structure of the l.h.s.\ of Eq.\ (\ref{eq:31})
in more detail.

The derivatives of $F(\hat\theta,\hat\lambda)$ with respect to
$\theta_j$ are calculated using
\begin{eqnarray}
  {\partial F(\hat\theta,\hat\lambda) \over \partial \theta_j} &=&
  \left(\sum_{n=1}^{N}
        {1 \over f(\hat\theta,\lambda_n)}
        {\partial f(\hat\theta,\lambda_n)\over\partial\theta_j}
  \right) F(\hat\theta,\hat\lambda), \label{eq:C3} \\
  {\partial^2 F(\hat\theta,\hat\lambda) \over
\partial \theta_j^2}
&=&
  \left(\sum_{n=1}^{N}
        {1 \over f(\hat\theta,\lambda_n)}
  {\partial^2 f(\hat\theta,\lambda_n)\over\partial\theta_j^2}
  \right) F(\hat\theta,\hat\lambda) +
 \nonumber \\ && \
  \left(\sum_{n \neq m}^{N}
  {1 \over f(\hat\theta,\lambda_n)
           f(\hat\theta,\lambda_m)}
  {\partial f(\hat\theta,\lambda_n)\over\partial\theta_j}
  {\partial f(\hat\theta,\lambda_m)\over\partial\theta_j}
  \right) F(\hat\theta,\hat\lambda). \label{eq:C4}
\end{eqnarray}
Inspection of Eqs.\ (\ref{eq:22}), (\ref{eq:C3}) and (\ref{eq:C4})
shows that
$\Delta_{\hat\theta} F(\hat\theta,\hat\lambda)$ has two
contributions, one involving a single summation over the channel
indices $n$, and another one involving a double summation over channel
indices $n \neq m$,
\begin{eqnarray} \label{gt}
  \Delta_{\hat\theta} F(\hat\theta,\hat\lambda) =
  \left( \vphantom{\sum_{n \neq m}^{N}}
    \sum_{n=1}^{N} g_1(\hat\theta,\lambda_n) \right)
    F(\hat\theta,\hat\lambda) +
  \left( \sum_{n \neq m}^{N}
g_2(\hat\theta,\lambda_n,\lambda_m)\right)
    F(\hat\theta,\hat\lambda). \label{eq:C5}
\end{eqnarray}
Using the definition (\ref{eq:22b}) of $f(\hat\theta,\hat\lambda)$ one may
straightforwardly calculate the functions $g_1$ and $g_2$. The expressions
are rather lengthy and will not be given here.

The r.h.s.\ of Eq.\ (\ref{eq:31}) contains the differential operator
$D_{\hat\lambda}$, which is given by Eq.\ (\ref{eq:29}). Simple
algebra yields
\begin{eqnarray}
  D_{\hat\lambda} &=& \sum_{n=1}^{N}
    \left( \lambda_n(1+\lambda_n) {\partial^2 \over \partial \lambda_n^2} +
           (1 + 2 \lambda_n) {\partial \over \partial \lambda_n} \right) +
  \nonumber \\ && \
  {\beta \over 2} \sum_{n \neq m}^{N} (\lambda_n - \lambda_m)^{-1}
    \left( \lambda_n(1+\lambda_n) {\partial \over \partial \lambda_n} -
           \lambda_m(1+\lambda_m) {\partial \over \partial
\lambda_m} \right).
\end{eqnarray}
As a consequence, $D_{\hat\lambda}
F(\hat\theta,\hat\lambda)$ has
again the
structure of Eq.\ (\ref{eq:C5}), with $g_1$
and $g_2$
now given by
\begin{mathletters} \label{gl}
\begin{eqnarray}
  g_1(\hat\theta,\lambda) &=&
    {1 \over f(\hat\theta,\lambda)}
    \left[\lambda (1+\lambda)
    {\partial^2 f(\hat\theta,\lambda) \over \partial \lambda^2} +
    (1+2\lambda){\partial f(\hat\theta,\lambda) \over
\partial \lambda}
    \right], \\
  g_2(\hat\theta,\lambda_1,\lambda_2) &=&
    {\beta/2 \over \lambda_1 - \lambda_2}
    \left[ {\lambda_1(1+\lambda_1) \over f(\hat\theta,\lambda_1)}
    {\partial f(\hat\theta,\lambda_1) \over \partial \lambda_1} -
    {\lambda_2(1+\lambda_2) \over f(\hat\theta,\lambda_2)}
    {\partial f(\hat\theta,\lambda_2) \over \partial
\lambda_2} \right].
\end{eqnarray}
\end{mathletters}%
Comparison of Eqs.\ (\ref{gt}) and (\ref{gl}) shows that the two definitions
of the functions $g_1$ and $g_2$ are identical. This completes the proof of
Eq.\ (\ref{eq:31}).

\section{Extension to higher dimensional supermatrices}
\label{app:d}

The argumentation presented in Secs.\ \ref{sec:3} and \ref{sec:4}
can be generalized to $\sigma$ models with $Q$ matrices
of arbitrary dimension $8p$ with $p \ge 1$. This generalized equivalence
proof applies to the $p$-point functions $\rho_p(T_1,\ldots,T_p)$ instead
to the limited number of
statistical quantities that can be generated by the ``small''
$\sigma$ model with $p=1$ (compare appendix \ref{app:b}). Here,
we briefly present the modifications with respect to the $p=1$ case.
The modifications concern the parameterization
(\ref{eq:Sigma}) and the generating function (\ref{eq:22b}).

The main technical difficulty in such a generalization is due to the
radial
part of the Laplace operator. The procedure to calculate it on conventional
symmetric spaces is standard \cite{helgason} and is carried over to the
supersymmetric $\sigma$ models as described in appendix B of Ref.\
\ref{mmz}. It is now more convenient to use a slightly modified form of
the parameterization of the $Q$-matrices, where $\hat\theta$ in Eq.\
(\ref{eq:6}) is fully diagonal (rather than block-diagonal):
\begin{eqnarray}
&& \hat\theta=
\left(\begin{array}{cc}
\hat x & 0 \\
0 &  i\hat y\\
\end{array}\right), \nonumber \\ && (\hat x)_{nm} = x_n \delta_{nm},
                    (\hat y)_{nm} = y_n \delta_{nm},
  1 \le n,m \le 2 p.
\label{eq:d1}
\end{eqnarray}
The symmetry restrictions are [cf.\ Eq.\ (\ref{eq:10})]
\begin{eqnarray}
\nonumber
y_i & = & y_{i+p} \hphantom{,\ y_i = y_{i+p}} \ \mbox{if $\beta=1$},\\
\label{eq:d2}
x_i &=& x_{i+p},\ y_i = y_{i+p} \ \mbox{if $\beta=2$},\\
\nonumber
x_i & = & x_{i+p} \hphantom{,\ y_i = y_{i+p}} \ \mbox{if $\beta=4$},
\end{eqnarray}
for $i=1,\ldots,p$. In the case $p=1$, we have the relations
$x_1=\theta_1+\theta_2$, $x_2=\theta_1-\theta_2$, $y_1=\theta_3+\theta_4$,
and $y_2=\theta_3-\theta_4$ between these parameters and the $\theta_i$
used in (\ref{eq:9}).

We can directly apply the results of appendix B in Ref.\ \ref{mmz}
which are given in terms of the so-called {\em roots \/} $\alpha(\Theta)$
[with $\Theta$ given as in (\ref{eq:a5})]. The roots are linear functions
of $\Theta$ which are the eigenvalues of the linear mapping
ad$(\Theta)(X_{\alpha}):=[\Theta,X_{\alpha}]=\alpha(\Theta) X_{\alpha}$,
defined on a certain super-Lie algebra.\cite{mmz} The eigenvectors $X_{\alpha}$
of the mapping are the {\em root vectors,} which do not depend on
$\Theta$. The radial integration measure $J(\hat\theta)$ in Eq.\ (\ref{eq:22})
can be expressed as\cite{helgason}
\begin{equation}
\label{eq:d3}
J(\hat\theta)=\prod_{\alpha>0} \sinh^{m_\alpha}
[{\case{1}{2}}\,\alpha(\Theta)],
\end{equation}
where the integer $m_\alpha$ is the multiplicity of the root $\alpha$ (the
dimension of the root space). Both positive and negative values of $m_{\alpha}$
can occur. The factor $\frac{1}{2}$ is due to the difference between the
normalization (\ref{eq:6}) of $\Theta$ and the one used in Ref.\
\ref{mmz}. In appendix A of Ref.\ \ref{mmz}, explicit formulas for the
roots as well as for the root vectors are given for the case $\beta=1$, $p=1$.

We have calculated the roots and the root vectors for all $\beta$
and arbitrary dimension $8p$ of the $Q$ matrices. For simplicity, we
only present the results for the roots and their multiplicities. Let us
denote with $p_x$ ($p_y$) the number of independent $x_i$ ($y_i$)
parameters, i.e.: $p_x=2p,p,p$ and $p_y=p,p,2p$ for $\beta=1,2,4$,
respectively.
Note that $\beta p_x = 2 p_y$. We find 8 different types of (positive) roots:
\begin{equation}
\label{eq:d4}
\begin{array}{lll}
\alpha(\Theta)=x_j-x_l \qquad & (1\le j<l\le p_x),
& \quad m_\alpha=\beta, \\
\alpha(\Theta)=x_j-i y_l \qquad & (1\le j\le p_x,\ 1\le l\le p_y),
& \quad m_\alpha=-2, \\
\alpha(\Theta)=i(y_j-y_l) \qquad & (1\le j<l\le p_y),
& \quad m_\alpha=4/\beta, \\
\alpha(\Theta)=2x_j \qquad & (1\le j\le p_x),
& \quad m_\alpha=\beta-1, \\
\alpha(\Theta)=2i y_j \qquad & (1\le j\le p_y),
& \quad m_\alpha=4/\beta-1, \\
\alpha(\Theta)=x_j+x_l \qquad & (1\le j<l\le p_x),
& \quad m_\alpha=\beta, \\
\alpha(\Theta)=x_j+i y_l \qquad & (1\le j\le p_x,\ 1\le l\le p_y),
& \quad m_\alpha=-2, \\
\alpha(\Theta)=i(y_j+y_l) \qquad & (1\le j<l\le p_y),
& \quad m_\alpha=4/\beta. \\
\end{array}
\end{equation}
The radial part of the Laplacian takes the form
\begin{equation}
\label{eq:d5}
\Delta_{\hat\theta} = \sum_{j=1}^{p_x}
J^{-1}(\hat\theta)\,{\partial \over \partial x_j}\,J(\hat\theta)\,{\partial
\over \partial x_j}+
{\beta \over 2} \sum_{j=1}^{p_y}
J^{-1}(\hat\theta)\,{\partial \over \partial y_j}\,J(\hat\theta)\,{\partial
\over \partial y_j}.
\end{equation}
The expressions (\ref{eq:23}) and (\ref{eq:7}) for the generating
function now remain valid with the modified $\hat\theta$ of Eq.\ (\ref{eq:d1}),
and with Eq. (\ref{eq:22b}) replaced by
\begin{equation}
\label{eq:d6}
f(\hat\theta,\lambda)=
{\displaystyle \prod_{i=1}^{p_y}[1+2\lambda+\cos(y_i)]}
{\displaystyle \prod_{i=1}^{p_x}[1+2\lambda+\cosh(x_i)]^{-\beta/2}}
{}.
\end{equation}
It is convenient to use the variables $u_i=\sinh^2(\frac{1}{2} x_i)$ and
$v_i=-\sin^2(\frac{1}{2} y_i)$ in terms of which the
Laplacian has the form
\begin{mathletters}
\begin{equation}
\label{eq:d7}
\Delta_{\hat\theta} = \sum_{j=1}^{p_x}
\tilde J^{-1}\,{\partial \over \partial u_j}\,u_j(1+u_j)\,
\tilde J\,{\partial \over \partial u_j}-
 {\beta \over 2} \sum_{j=1}^{p_y}
\tilde J^{-1}\,{\partial \over \partial v_j}\,v_j(1+v_j)\,
\tilde J\,{\partial \over \partial v_j},
\end{equation}
\begin{eqnarray}
\tilde J & = &
\prod_{1\le i<j\le p_x} (u_i-u_j)^\beta
\prod_{1\le i<j\le p_y} (v_i-v_j)^{4/\beta}
\prod_{i=1}^{p_x}\prod_{j=1}^{p_y} (u_i-v_j)^{-2}
 \\ \nonumber && \times
\prod_{i=1}^{p_x}\Bigl(u_i(1+u_i)\Bigr)^{\beta/2-1}
\prod_{i=1}^{p_y}\Bigl(v_i(1+v_i)\Bigr)^{2/\beta-1}.
\label{eq:d8}
\end{eqnarray}
\end{mathletters}%
The generating function $F(\hat\theta,\hat\lambda)$ is given by
\begin{equation}
\label{eq:d9}
F(\hat\theta,\hat\lambda)=
\prod_{n=1}^{N}
\left(
{\displaystyle \prod_{i=1}^{p_y}(1+\lambda_n+v_i)}
{\displaystyle \prod_{i=1}^{p_x}(1+\lambda_n+u_i)^{-\beta/2}}
\right).
\end{equation}

We have verified that the identity of Laplacians [Eq.\ \ref{eq:CentralResult})]
remains true for the modified expressions (\ref{eq:d7}) and (\ref{eq:d9}).
The calculations goes in a similar way as shown in App.\ \ref{app:c} for
$p=1$. Now, we have to keep track of 7 different types of contributions
with double and triple sums over functions of $\lambda_n$, $u_i$, $v_j$.

In App.\ \ref{app:b} we have shown that the average density
of transmission eigenvalues $\rho(T)$ can be
obtained from the generating function (\ref{eq:23}). Using the
corresponding function for the higher-dimensional $\sigma$ model
considered here, it is straightforward to get the $p$-point correlation
functions $\rho_p(T_1,\ldots,T_p)$.

\end{document}